\documentclass{webofc}
\usepackage[varg]{txfonts}  
\usepackage{graphicx,epsfig,amsmath,amssymb,amsbsy,natbib}
\woctitle{Physics at the magnetospheric boundary}
\begin{document}
\title{Magnetic field structure in accretion columns on HMXB and effects on CRSF}

\author{Dipanjan Mukherjee\inst{1}\fnsep\thanks{\email{dipanjan@iucaa.erner.in
        }} \and
        Dipankar Bhattacharya\inst{1}\fnsep
        \and
        Andrea Mignone \inst{2}\fnsep
        % etc.
}

\institute{Inter University Center for Astronomy and Astrophysics, post bag 4, Ganeshkhind, Pune, India. 
\and
Dipartimento di Fisica Generale, Universita di Torino, via Pietro Giuria 1, 10125 Torino, Italy.
          }

\abstract{%
In accreting neutron star binaries, matter is channelled by the magnetic fields from the accretion disc to the poles of neutron stars forming an accretion mound. We model such mounds by numerically solving the Grad-Shafranov equation for axisymmetric static MHD equilibria. From our solutions we infer local distortion of field lines  due to the weight of accreted matter. Variation in mass loading at the accretion disc will alter the shape of the accretion mound which will also affect the local field distortion. From simulations of cyclotron resonance scattering features from HMXBs, we conclude that local field distortion will greatly affect the shape and nature of the CRSF. From phase resolved spectral analysis one can infer the local field structure and hence the nature of mass loading of field lines at the accretion disc. We also study the stability of such mounds by performing MHD simulations using the PLUTO MHD code. We find that pressure and gravity driven instabilities depend on the total mass accreted and the nature of mass loading of the field lines. 
}
\maketitle
\section{Introduction}
\label{intro}
In accreting neutron stars with an extended magnetosphere matter proceeds along the fields towards the magnetic poles. The infalling matter slows down and settles at the base to form an accretion mound confined by the magnetic field.  Local magnetic field will be distorted by the pressure of the accreted matter. As mounds of larger mass are built, leading to larger field distortions, MHD instabilities may be triggered \cite{litwin01} which will cause matter to escape from the confined mound. The magnetic structure inside the accretion column and its local dynamics will have important concequences on the observed X-ray emission from such system. For neutron stars with high surface magnetic field ($\sim 10^{12}$), the X-ray emission show cyclotron scattering features which depend on the local magnetic field, and hence can be used as a tool to probe the structure of the accretion column. 

For the work in this presentation we consider a neutron star (mass $\sim 1.4 M_{\bigodot}$ and radius $\sim 10$ km) with surface field ($\sim 10^{12}$ G) with an accretion of radius $\sim 1$km. We have studied the structure of the axisymmetric accretion mound by recasting the time-independent Euler equation as the Grad-Shafranov (hereafter GS) equation. We solve the GS equation for accretion mounds of different shapes and sizes and discuss the effect of mass loading of the field on the shape of the mound. We show from MHD simulations that stable mounds can be built only up to a threshold height, beyond which MHD instabilities are triggered. We discuss the effect of the field distortion and local dynamics on the emitted CRSF spectra.  

\section{Grad-Shafranov solutions for accretion mounds}\label{sec.GS}
\begin{figure}
\centering
\includegraphics[width = 4.5cm, height = 4.5cm,keepaspectratio] {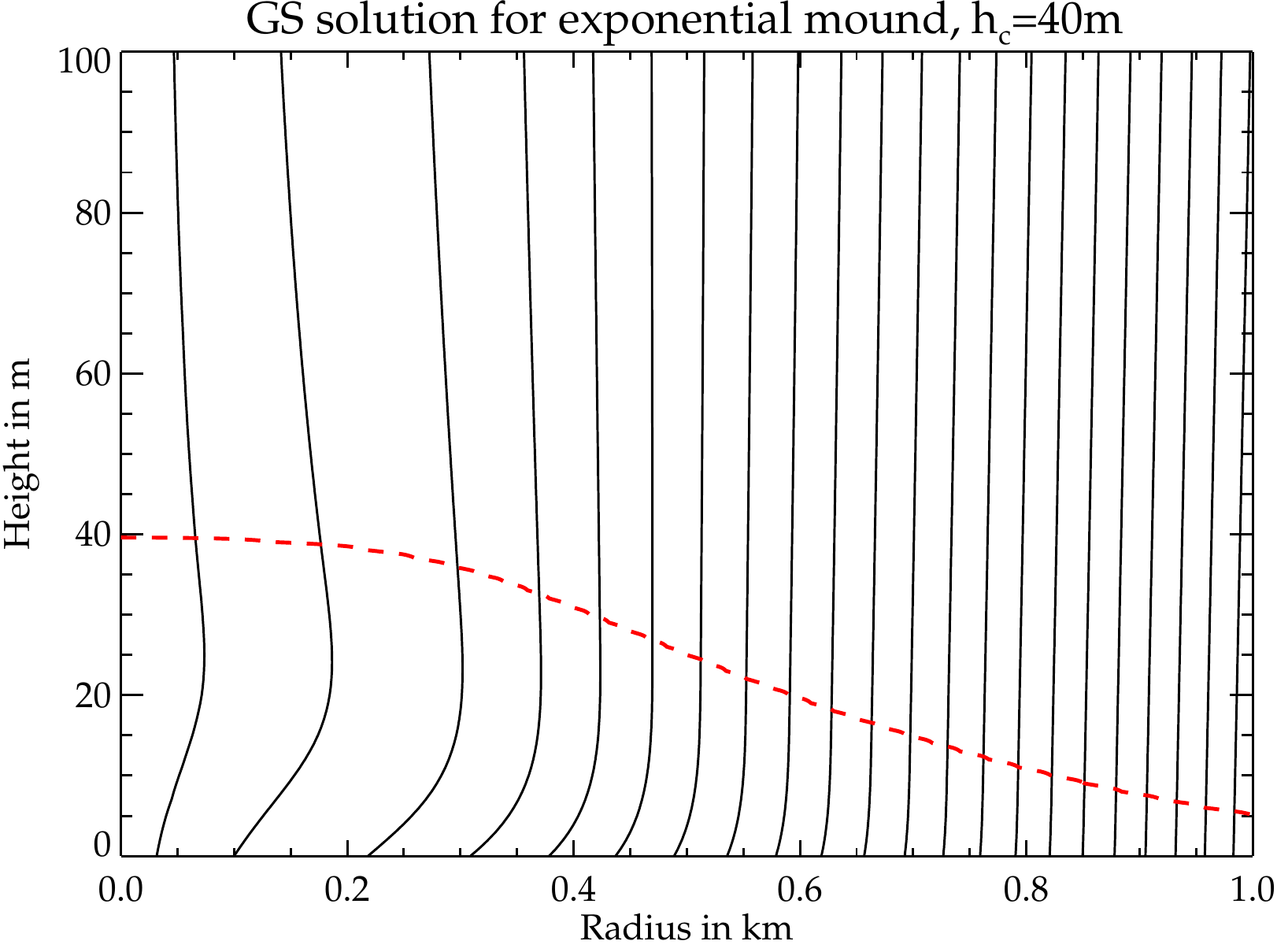}
\includegraphics[width = 4.5cm, height = 4.5cm,keepaspectratio] {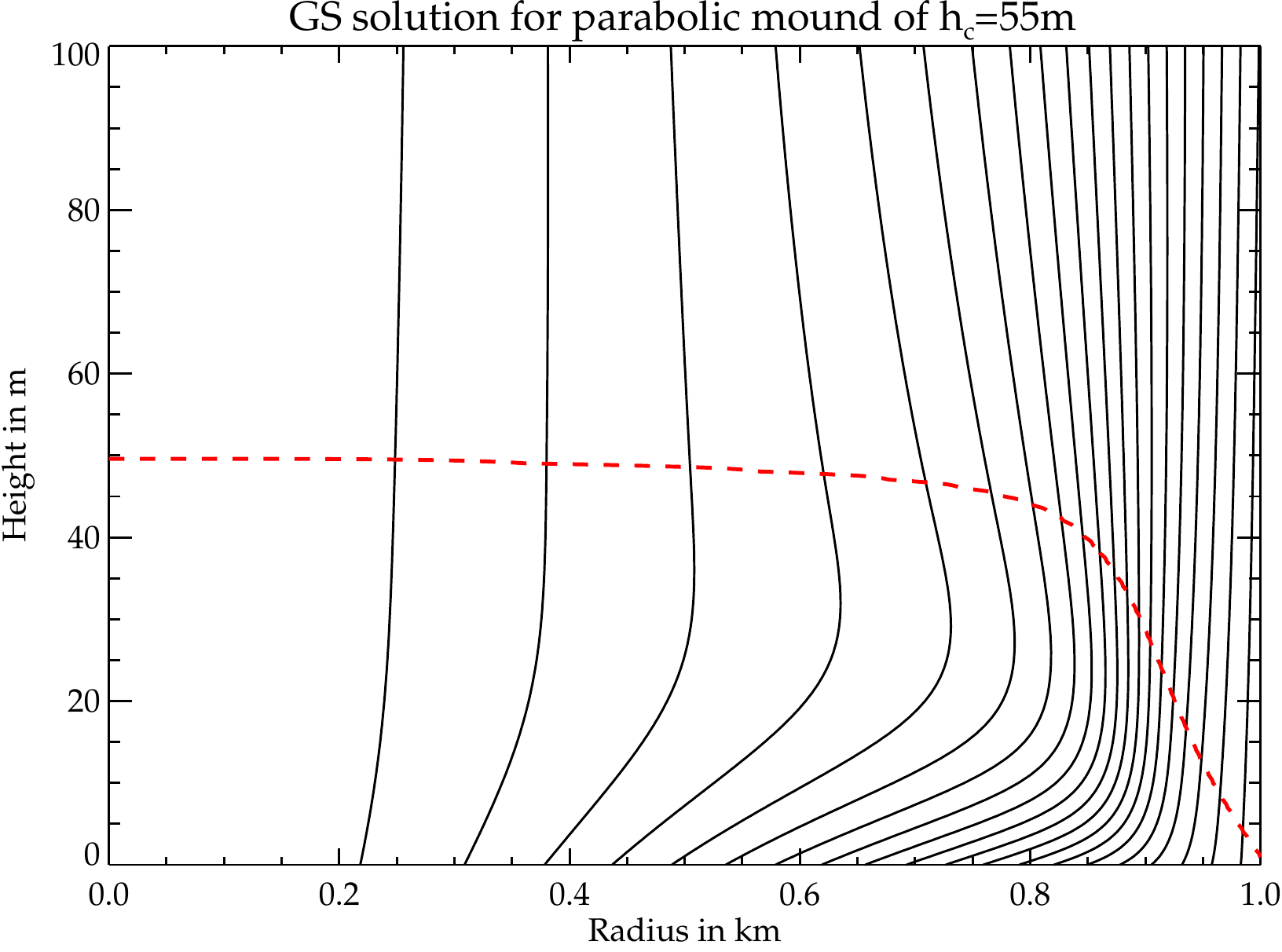}
\includegraphics[width = 4.5cm, height = 4.5cm,keepaspectratio] {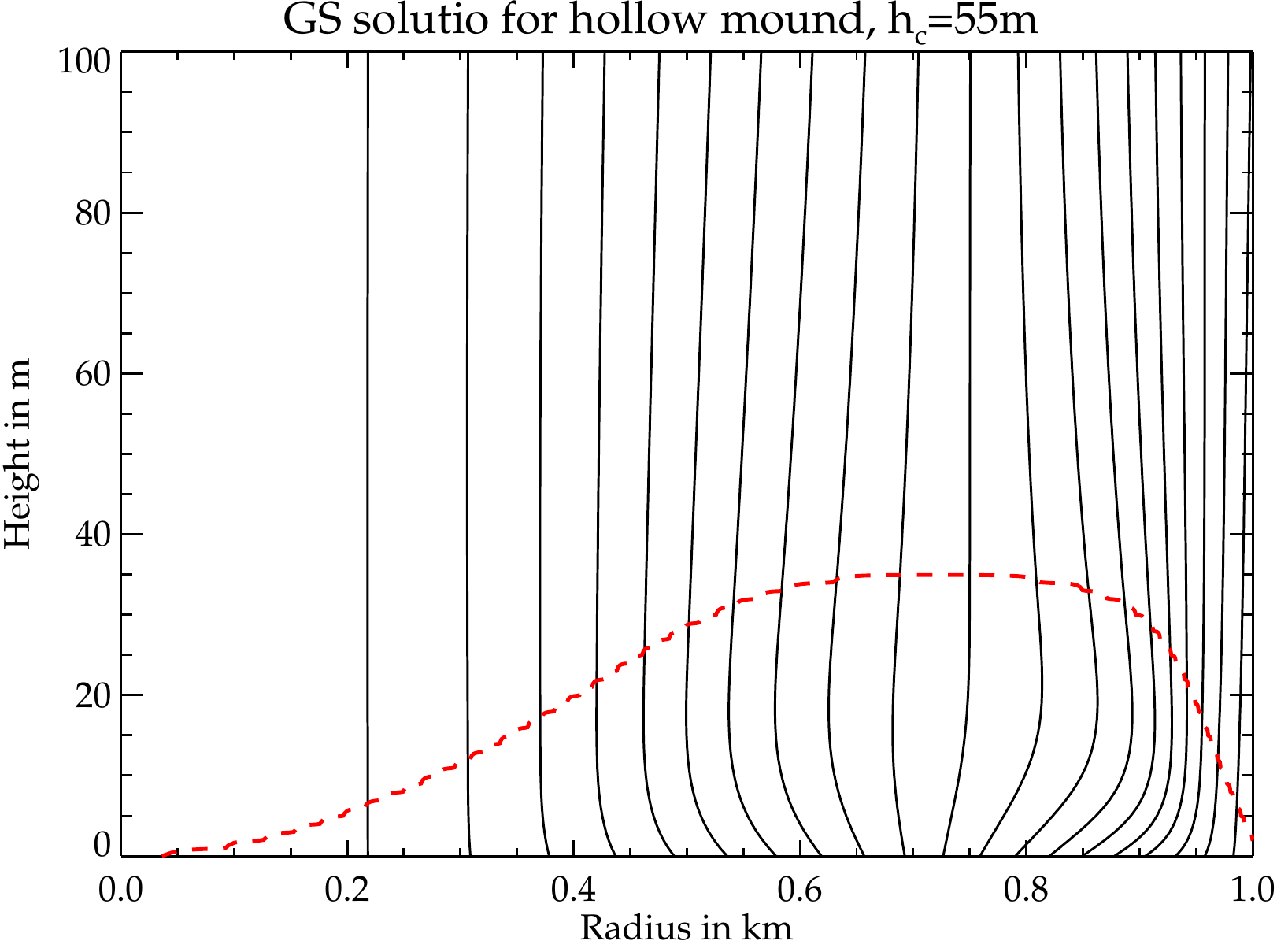}
\includegraphics[width = 4.5cm, height = 4.5cm,keepaspectratio] {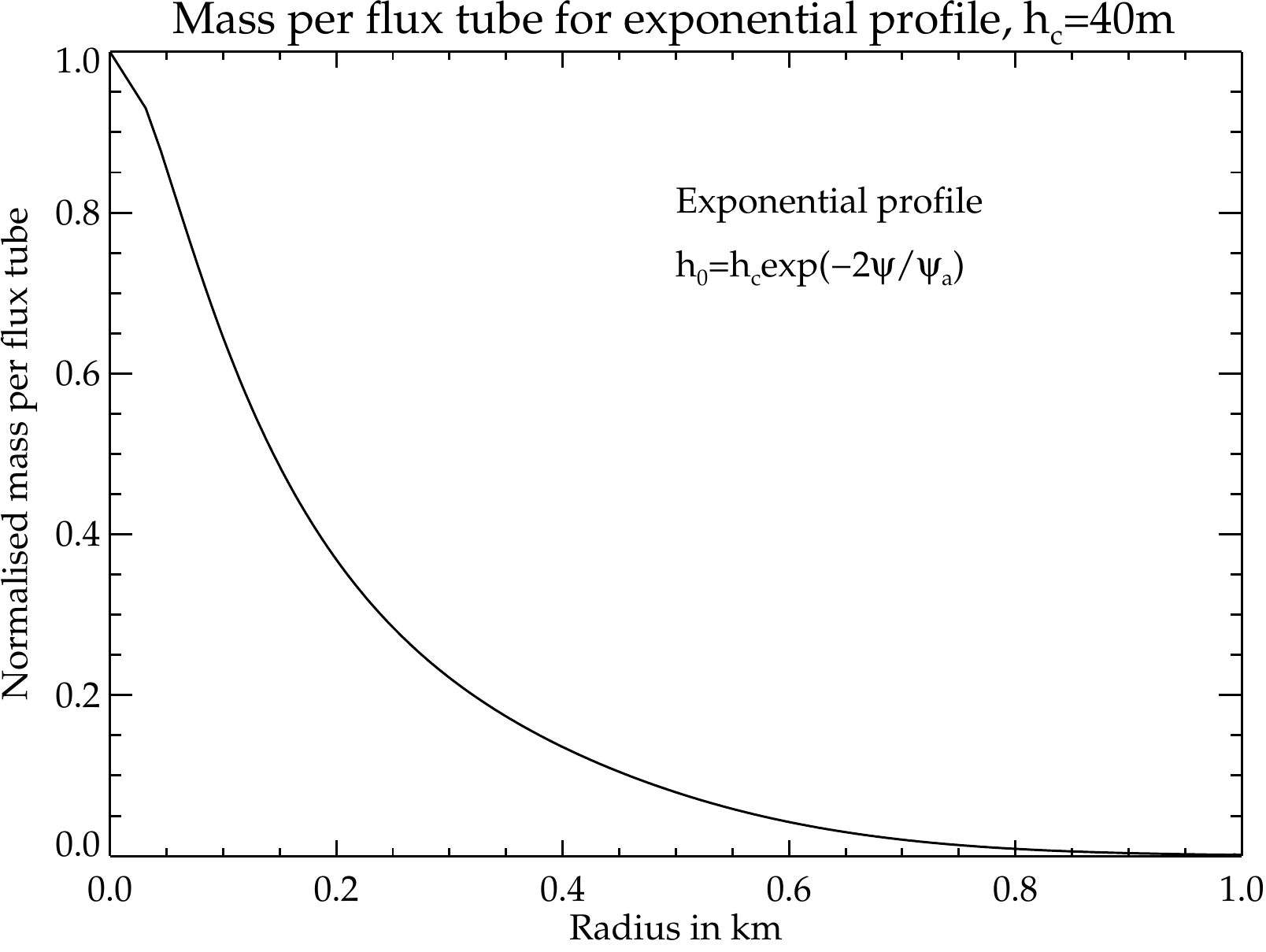}
\includegraphics[width = 4.5cm, height = 4.5cm,keepaspectratio] {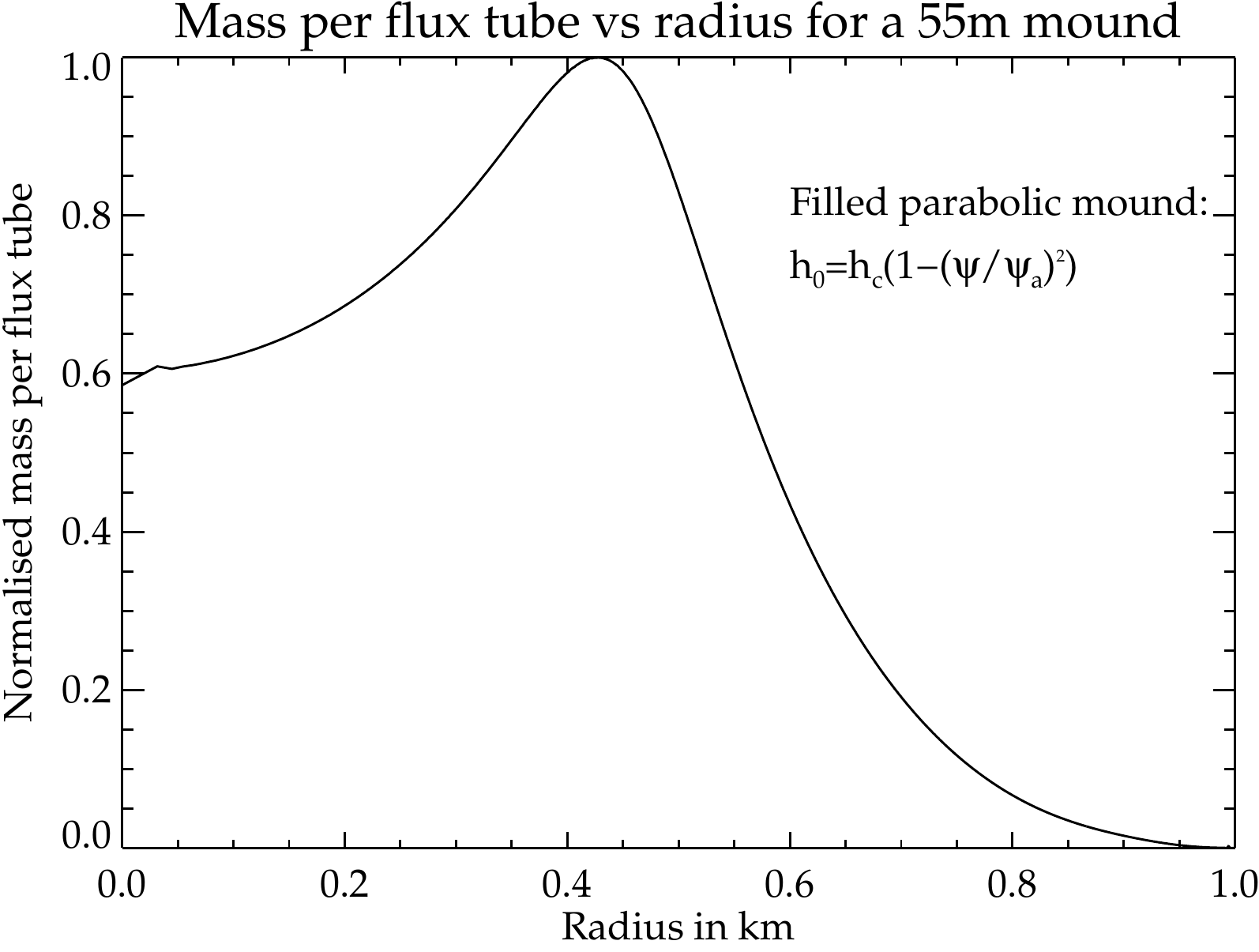}
\includegraphics[width = 4.5cm, height = 4.5cm,keepaspectratio] {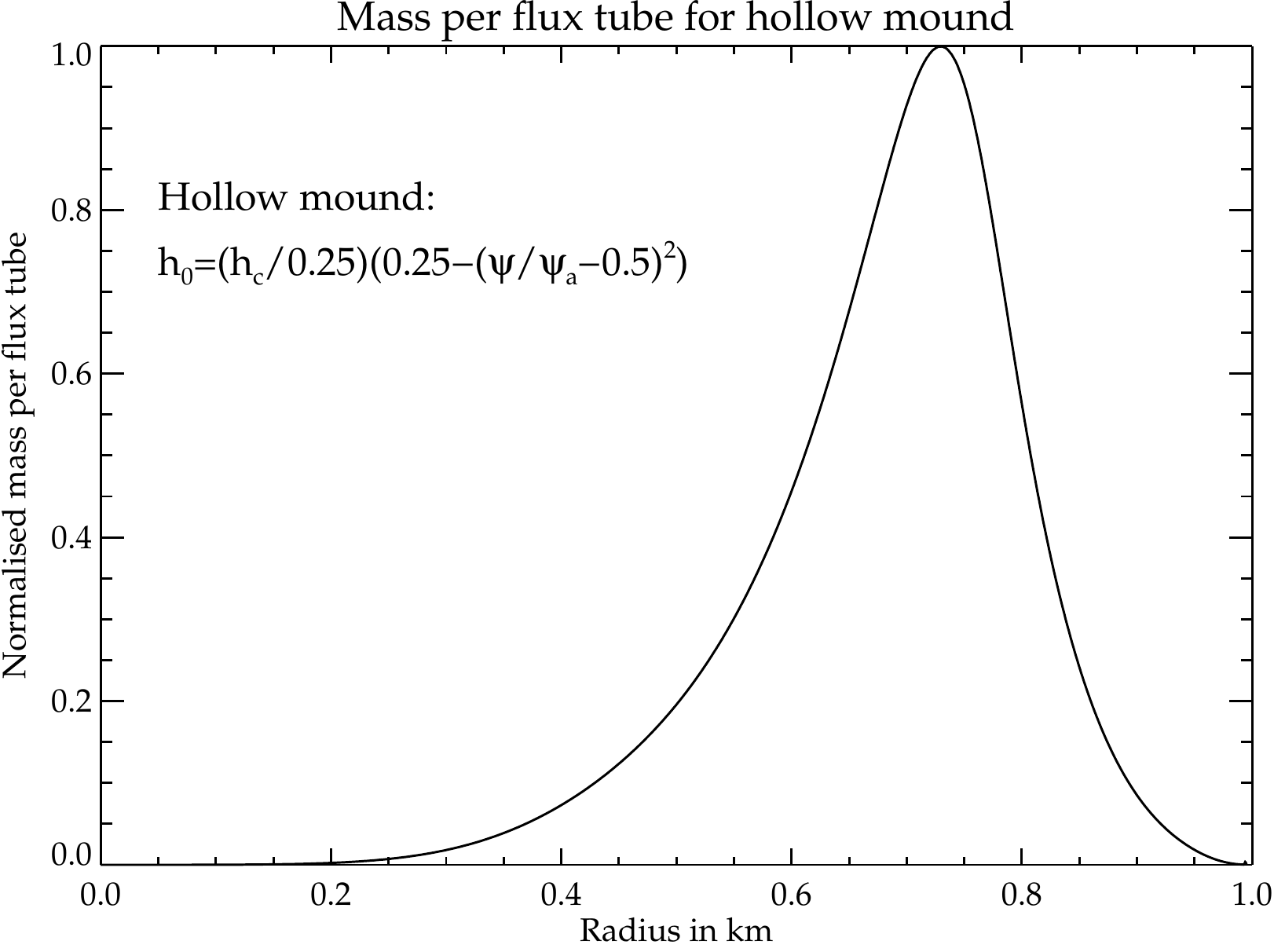}
\caption{\small Top: GS solutions for exponential $\left(h_0=h_c\exp{\left(-2\psi/\psi _a\right)}\right .$, $\psi _a$ being the flux function at $r=1$km), parabolic $\left( h_0=h_c\left(1-\left(\psi/\psi _a\right)^2 \right)\right)$ and hollow $\left(h_0=h_c/0.25\left(0.25-\left(\psi/\psi _a - 0.5\right)^2\right)\right)$ mound profiles. The red dotted line represents the top of the mound. Bottom: Mass per flux tube vs radius for respective mound profiles.}
\label{GSsol}       
\end{figure}

\begin{figure}
\centering
\includegraphics[width = 12cm, height = 12cm,keepaspectratio] {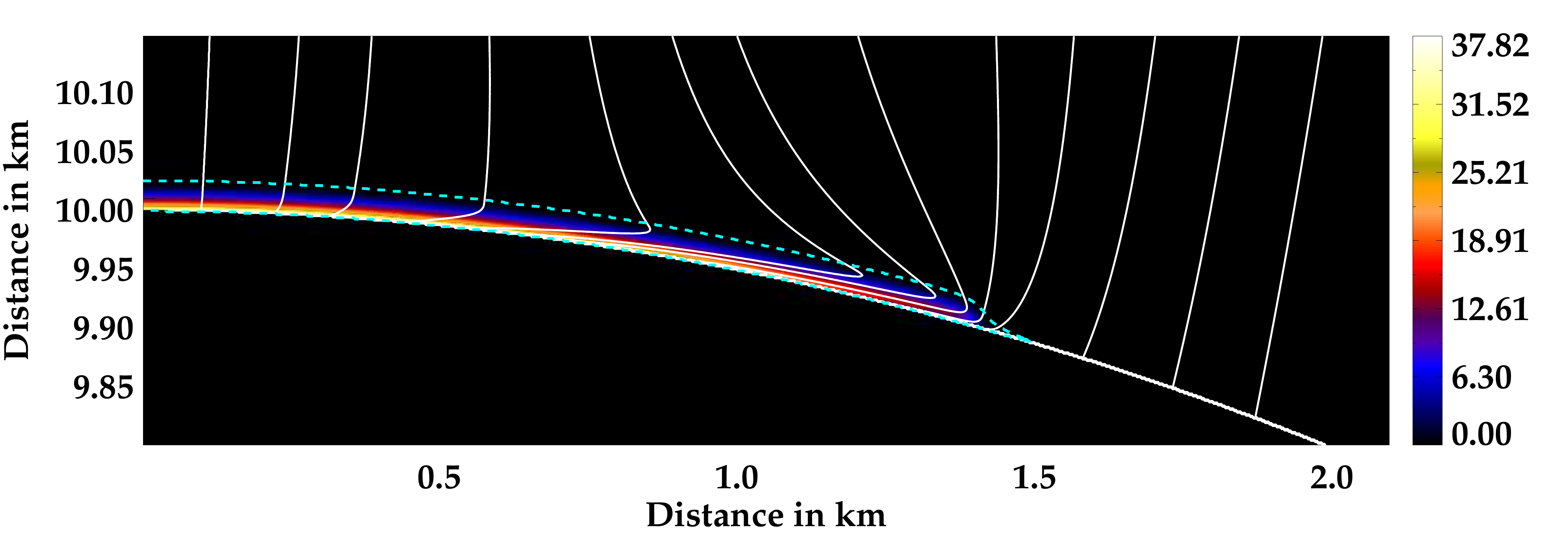}
\caption{\small GS solution for a mound of height $h_c=21$ m with a polar cap radius $R_p\sim 1.5$ km with surface field $B_p=10^{11}$G. The white lines depict the magnetic field. The density is represented in coloured contours, with the blue-dotted line marking the top of the mound.}
\label{GSsph}       
\end{figure}

Following the formalism of Mukherjee et.~al \cite{mukherjee12}, \cite{mukherjee13a}, \cite{mukherjee13b}, \cite{mukherjee13c} we model the accretion mound by solving for the static Euler equation. Assuming axisymmetry one can express the poloidal magnetic field in terms of the flux function ($\psi$) to write the Grad-Shafranov (GS) equation for the static accretion mound in force equilibria:
\begin{eqnarray}
r\frac{\partial}{\partial r}\left(\frac{1}{r}\frac{\partial \psi}{\partial r}\right) + \frac{\partial ^2 \psi}{\partial z^2} &=& -4 \pi r^2 \rho g \frac{dh_0(\psi)}{d\psi} \quad \quad \quad \, \mbox{ (Cylindrical coordinates)} \label{eq.gs.cyl} \\
\frac{\partial ^2 \psi}{\partial r^2}+\frac{\sin \theta}{r^2} \frac{\partial }{\partial \theta}\left(\frac{1}{\sin \theta}\frac{\partial \psi}{\partial \theta}\right) &=& -4 \pi r^2 \sin ^2 \theta \rho g \frac{d h_0(\psi)}{d \psi} \quad \mbox{(Spherical coordinates)} \label{eq.gs.sph}
\end{eqnarray}
where $\rho$ is the density and $g=1.86\times10^{14} \mbox{ g cm}^{-2}$ is the acceleration due to gravity. The mound height function $h_0(\psi)$ is an arbitrary function of the magnetic flux ($\psi$) which defines the shape of the mound. We have used the Paczynski equation of state \cite{pacz83} \cite{mukherjee13b} $p= (8 \pi/15)m_ec^2\left(\frac{m_ec}{h}\right)^3 x_F^5/\left(1+16/25x_F^2\right)^{1/2}$, which closely approximates (to within 1.5\%) the $T=0$ degenerate Fermi equation of state for electrons. The density is obtained from the Fermi momentum ($x_F=\frac{1}{m_e c} \left(\frac{3 h^3}{8 \pi \mu _e m_p}\right)^{1/3} \rho ^{1/3}$) as: 
\begin{equation}\label{fermiP}
x_F = \frac{5}{4}\left( \frac{\xi ^2 - 8/3 + \xi \sqrt{16/9 + \xi ^2}}{32/9} \right) ^{1/2} \; ; \; \xi=\frac{16}{15} \frac{\mu _e m_p}{m_e c^2} \left(h_0(\psi) - h\right) + 1 \end{equation}
The variable $h$ denotes the vertical height, which for cylindrical system is the coordinate $z$ and $r$ for spherical polar coordinates, such that $h_0-h$ is the height above the neutron star surface. The structure of accretion mounds on high field pulsars have been previously studied by solving the GS equation in cylindrical coordinates (eq.~\ref{eq.gs.cyl}) \cite{mukherjee12}. Although realistic mound profiles can be of complex shapes depending on the mass-loading of the field lines by the accretion disc (or mass capture from winds), in this work we consider smoothly varying mound height profiles (e.g. $h(\psi)=h_c\left(1-\left(\psi/\psi_a\right)^2\right)$) for ease in numerical computations. The shape of the mound will however depend on the mass distribution in different flux-tubes. For a cylindrical coordinate system, the mass per flux tube is approximately related to the accretion rate per unit area as (following eq.~14, eq.~15 and eq.~17 of Hameury (1983) \cite{hameury83}):
\begin{equation}\label{eq.mdot}
\dot{m}t\simeq B_0 \int ^{h_0(\psi)}_{0} \rho(\psi _0,z) r(\psi_0) \frac{\partial r(\psi _0)}{\partial \psi} dz
\end{equation}
where $B_0=10^{12}$ G is the scale magnetic field. Integrating eq.~(\ref{eq.mdot}) using the GS solutions obtained previously we get the mass loading profile of a given mound (see Fig.~\ref{GSsol}). For the mound with exponential profile, the maximum mass is concentrated in flux tubes near the axis. For filled parabolic and hollow mounds, the maximum mass is in the flux tubes near the middle of the mound. In real systems a similar behaviour is expected as the mass-loading will be effective over a finite range of radii at the accretion disc with a maxima near the middle. Realistic models of accretion mounds should consider mass loading profiles from numerical studies of accretion funnel flows \cite{ghosh78}, \cite{romanova02} to determine the mound height function.
The solutions discussed so far were for neutron stars with surface fields $\sim 10^{12}$G with polar cap radius 1 km. To study the effect of low field pulsars with larger polar cap, we have solved the GS equation in spherical coordinates (eq.~\ref{eq.gs.sph}). A typical solution for a neutron star with surface field $10^{12}$G and $\sim 1.5$ km polar cap is presented in Fig.~\ref{GSsph}. We have repeated the analysis of Mukherjee et.~al 2012 \cite{mukherjee12} to evaluate the GS equation for mounds with different base magnetic fields, and now also for different polar cap radii. As reported earlier in Mukherjee et.~al 2012 \cite{mukherjee12} the GS solver does not converge beyond a threshold height for a given surface field. Additionally, we find that the threshold height also depends on polar cap size.  From numerical fits we find that threshold heights depends on the surface field and the polar cap radius (for a parabolic profile) as 
\begin{equation}\label{eq.threshold}
h_T=54\mbox{ m}\left(\frac{B}{10^{12}{\rm G}}\right)^{0.41} \left(\frac{R_p}{1{\rm km}}\right)^{0.42}
\end{equation}
The above scaling can be understood by considering the magnetostatic equilibrium between the pressure gradients and the tension from the curved field line: $ \frac{\partial p}{\partial r} \simeq B_z \frac{\partial B_r}{\partial z}$. Approximating the spatial derivatives with their length scales in either direction, and assuming that the radial component of the field is approximately related to the vertical component as $B_r=B_zR_p/h_T$ ($h_T$ being the threshold mound height), we get $p \simeq B_z^2 R_p^2/h_T^2$. The solutions do not converge when the local pressure gradients are larger than the tension from field curvature, resulting in the formation of closed magnetic loops. As the regions of maximum magnetic tension are near the middle of the mound where $x_F \sim 1$, the pressure can be approximated as $p\propto \rho ^{5/3} \propto h_T^{5/2}$, where the last relation follows from the approximation $x_F << 1$ in eq.~\ref{fermiP}. Using the above relations we get the threshold height as $h_T \propto B_z^{4/9}R_p^{4/9}$. The exponents thus obtained are very close to the values obtained from numerical fits (in eq.~\ref{eq.threshold})\footnote{The approximate analysis in Mukherjee et.~al (2012) \cite{mukherjee12} is similar in nature. However, effects due to bending of the field lines were not accounted for properly, as has been done in Titos et.~al \cite{titos13}, following which we get a better agreement between approximate exponents and the values obtained numerically}.

\section{MHD instabilities in the mounds}
\begin{figure}
\centering
\includegraphics[width = 12cm, height = 12cm,keepaspectratio] {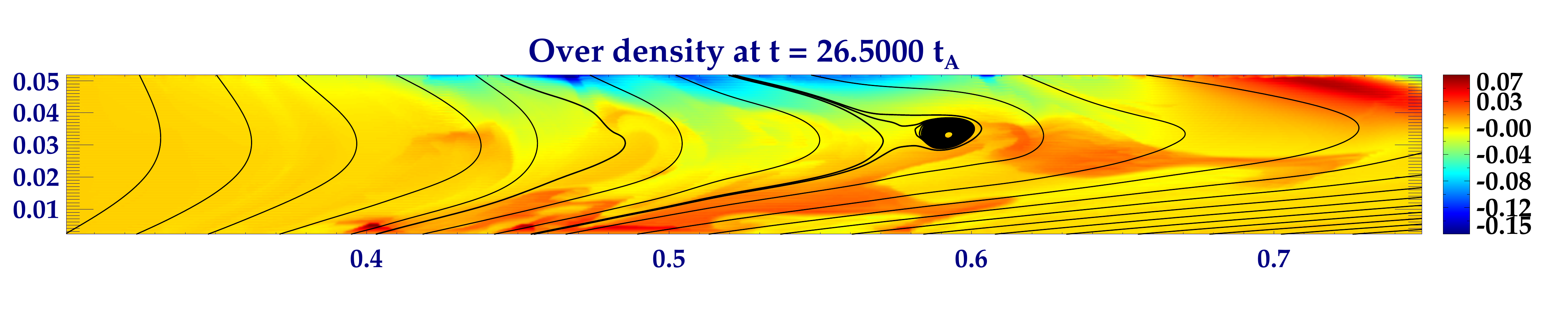}
\caption{\small Closed magnetic loops in 2D simulations for a $h_c=65$m mound with surface field $10^{12}$G. As the added mass descends due to gravity, it drags the field lines resulting in the formation of closed magnetic bubbles.}
\label{instability2D}       
\end{figure}
\begin{figure}
\centering
\includegraphics[width = 8cm, height = 8cm,keepaspectratio] {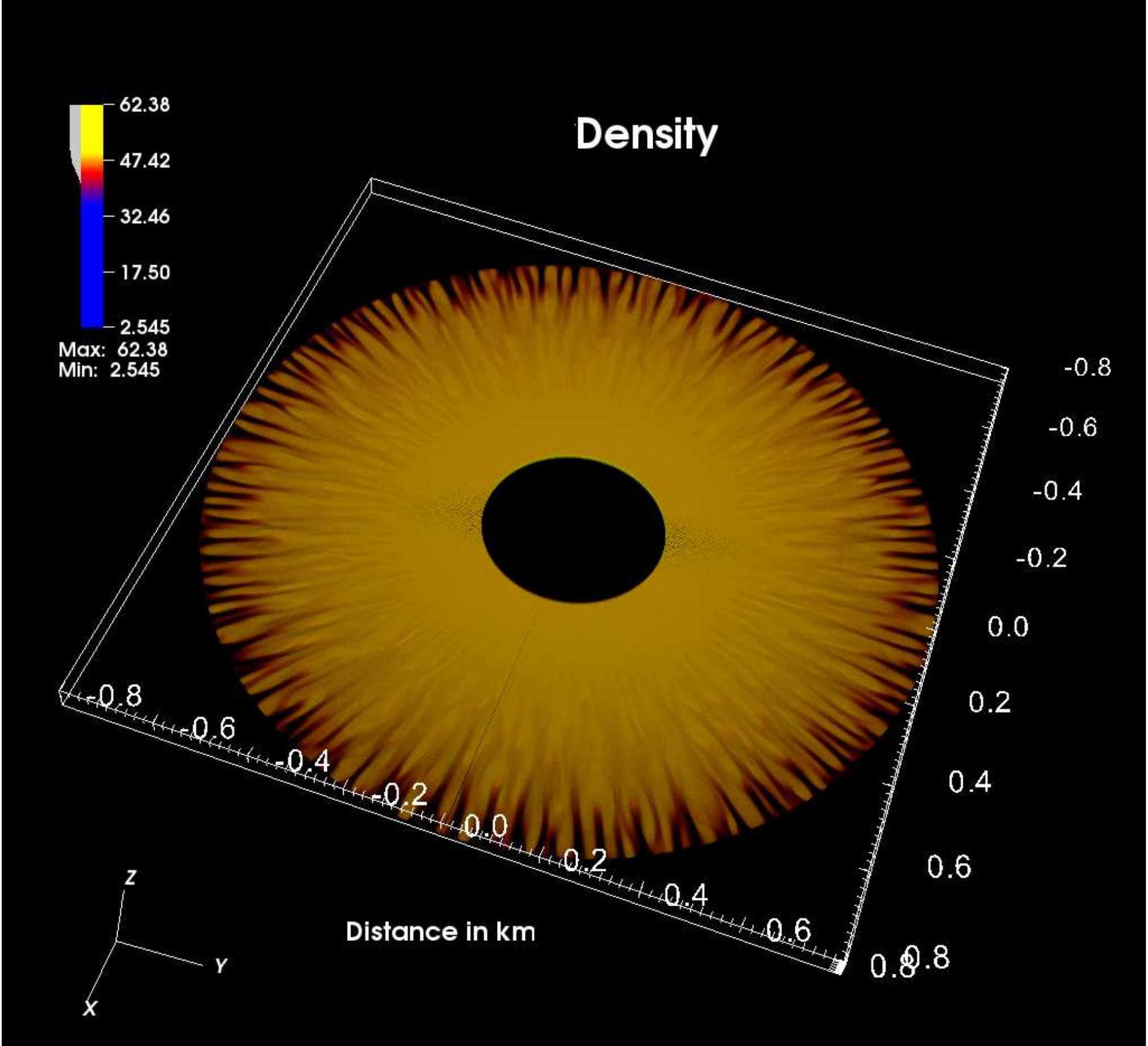}
\caption{\small 3D density contour of a perturbed mound of height $h_c=70$m at $t\sim 1 t_A$. The radial finger-like structures along the toroidal direction are due to pressure driven interchange instabilities.}
\label{instability3D}       
\end{figure}
The high $\beta$ plasma inside the mound along with highly curved magnetic field structures are prone to interchange and ballooning type pressure driven instabilities. MHD simulations of perturbed GS solutions confirm the presence of MHD instabilities beyond a threshold mound \cite{mukherjee13a} \cite{mukherjee13c}. 2D axisymmetric simulations \cite{mukherjee13a} show that gravity driven modes cause formation of closed loops inside mounds beyond a threshold size as added mass descends (as in Fig.~\ref{instability2D}). The threshold height from 2D simulations corresponds to the threshold of the GS solutions (as discussed in Sec.~\ref{sec.GS}) due to the non-convergence of the solutions. In 3D simulations, interchange type pressure driven instabilities take precedence \cite{mukherjee13c}. Multiple finger like channels are formed along the toroidal direction at the outer radial edges (see Fig.~\ref{instability3D}). Matter streams settles in magnetic valleys and flows outwards radially. We find a new threshold mound size beyond which MHD instabilities operate, lower than that of 2D simulations. For system with surface polar field $\sim 10^{12}$G, the instabilities are triggered for mounds of size larger than $5 \times 10^{-13} M_{\bigodot}$, which is in rough agreement to the instability threshold previously predicted by Litwin et.~al \cite{litwin01}. The onset of instabilities will have important concequences for the local dynamics. Leakage of matter on addition of mass beyond the instability threshold will limit further deformation of field lines. Also, radial outflows from the base of the `leaky column' will result in the spread of hot plasma on the surface which may manifest itself as an increase in the area of black body radiation from the surface of the star as is seen in RX J0440.9+4431 \cite{carlo13}.

\section{Effects on CRSF}
Scattering in the presence of strong magnetic field results in the formation of cyclotron resonance scattering features (CRSF), which appear as absorption like features in the X-ray spectra from high field neutron stars. The line energy of the CRSF depends on the local magnetic field:
\begin{equation}\label{cyclotronE}
E_n = m c^2 \frac{\left(1+2n(B/B_{\rm crit})\sin ^2 \theta\right)^{1/2} -1}{\sin ^2 \theta} \frac{1}{1+z}
\end{equation}
where $B_{\rm crit}=(m^2 c^3)/(e \hbar)$, $\theta$ is the angle between the scattering particle and the local magnetic field and $z=1/(\sqrt{1-(2GM)/(Rc^2)})-1$ is the gravitational red-shift for a neutron star of mass $M$ and radius $R$. Accretion induced distortions in the local field will be imprinted on to the emergent spectra. Mukherjee et.~al \cite{mukherjee12} \cite{mukherjee13a} have shown that field distortion can give rise to complex asymmetric profiles with multiple dip-like features by approximately evaluating the line profiles integrated over the mound. Overlying accretion column in accreting system will mask the effect of the field distortion, but if the emission region comes closer to the mound surface effects of field distortion will become apparent. Indeed, larger equivalent widths of the CRSF in V\,0332+53 for spectra closer to the surface (at lower luminosities) \cite{tsygan10} along with the reported asymmetry of the line profile \cite{katja05} \cite{nakajima10} hints to the possibility that the region of emission may have complex non-dipolar field structure. 

\begin{figure}
\centering
\includegraphics[width = 7cm, height = 7cm,keepaspectratio] {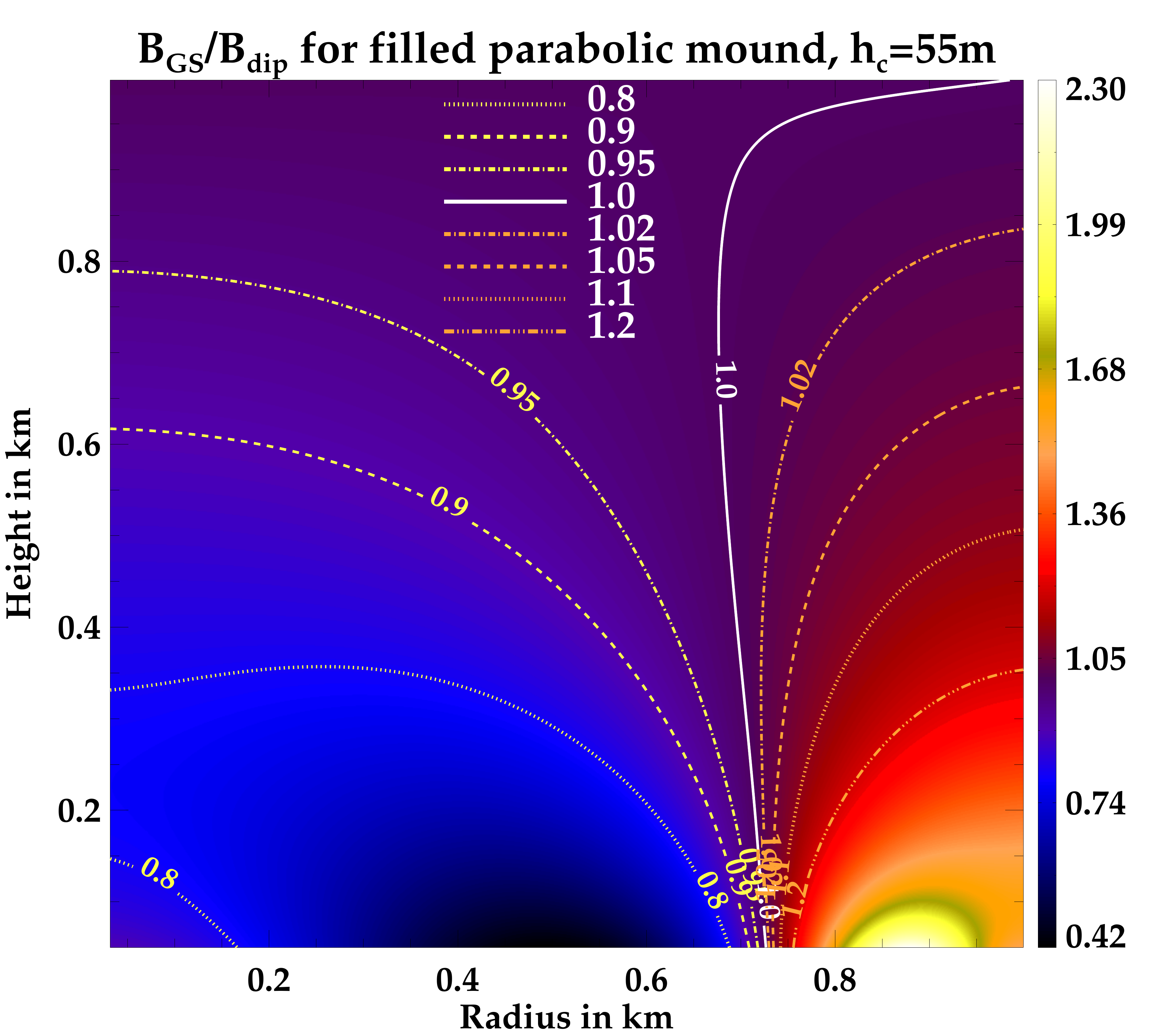}
\includegraphics[width = 7cm, height = 7cm,keepaspectratio] {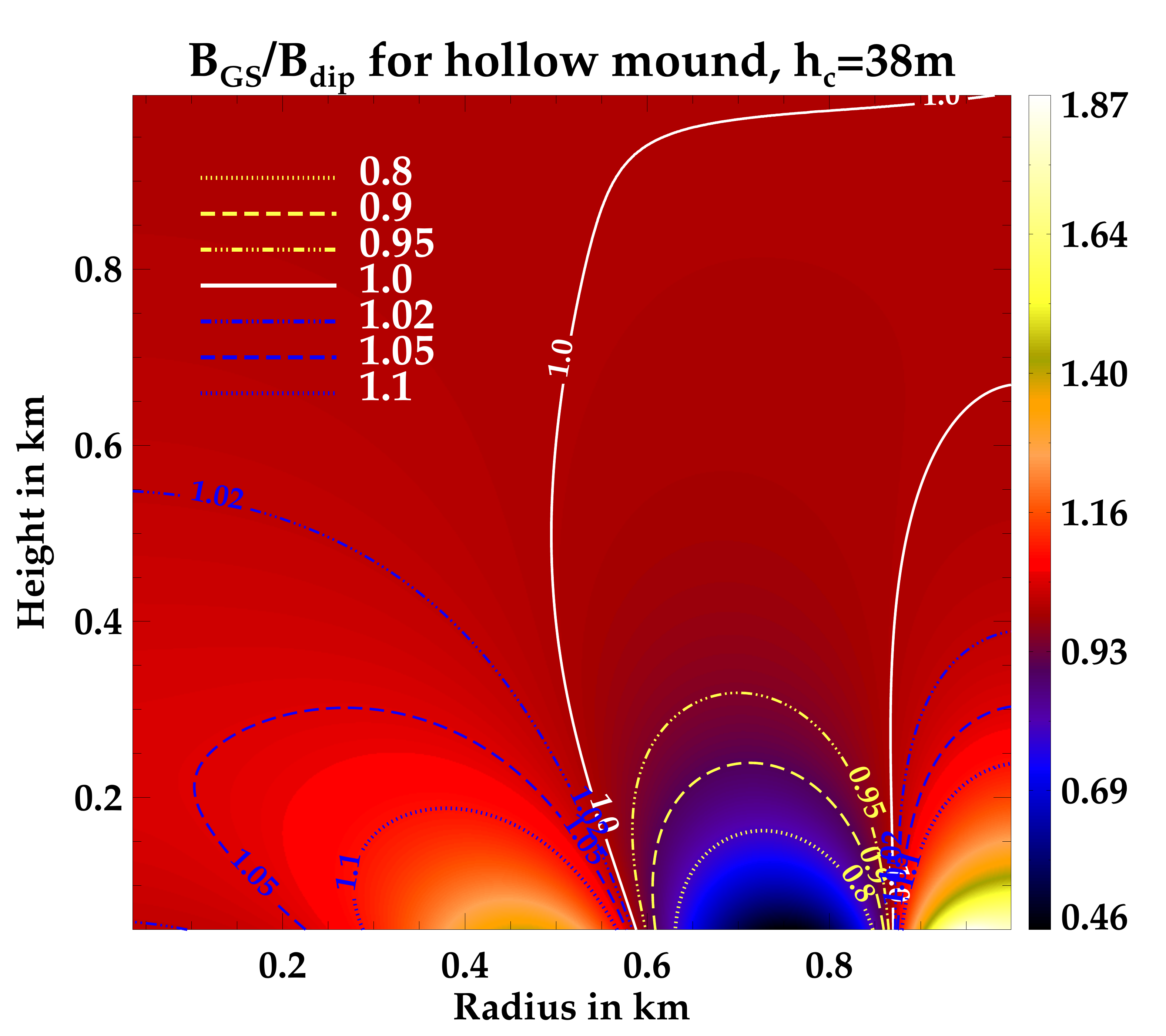}
\caption{\small Left: Ratio of field from GS solution to dipole value ($B_{\rm GS}/B_{\rm dip}$) for a filled parabolic mound of central height $h_c = 55$m. Even for heights $\sim 500$m, the GS solution shows $\sim 5\%$ deviation from dipole value. Right: $B_{\rm GS}/B_{\rm dip}$ for a hollow mound of apex height 38m.}
\label{Bcompare}       
\end{figure}

From the current GS solutions we find that  for heights below 200-300 m, there is significant distortion in the local field (see Fig.~\ref{Bcompare}), which can affect the CRSF spectra. As magnetic field is pushed outwards by pressure of accreted matter in filled mounds, magnetic field is larger than dipole value for $r \gtrsim 700$ m. For hollow mounds, field is enhanced on both sides as it is pushed outwards on both the inner and outer edges , lowering in value  in the central region near the apex of the mound. If the CRSF is originating from regions of distorted fields, the inferred magnetic field will not reflect the dipole moment of the neutron star, as is commonly associated in the current studies so far. Comparison of magnetic field obtained from CRSF and that from tracking the Alfv\'en radius through disc-magnetosphere interaction or QPO studies etc. will show the extent of local non-dipolar nature of the magnetic field.

%\bibliography{newbib}
\def\apj{ApJ}%
\def\mnras{MNRAS}%
\def\aap{A\&A}%
\def\apjl{ApJ}
\def\physrep{PhR}
\def\apjs{ApJS}
\def\pasa{PASA}
\def\pasj{PASJ}
\def\nat{Nature}
\def\memsai{MmSAI}

\end{document}